\documentclass[%
 reprint,
superscriptaddress,
 amsmath,amssymb,
 aps,
]{revtex4-2}

\usepackage{graphicx}
\usepackage{dcolumn}
\usepackage{bm}

\usepackage{xcolor}

\begin{document}


\title{Is a model equivalent to its computer implementation?}

\author{Beatrix C. Hiesmayr}
\affiliation{Universit\"at Wien, Fakult\"at f\"ur Physik, W\"ahringerstrasse 17, Vienna, Austria}
\author{Marc-Thorsten H\"utt}
\affiliation{School of Science,  Constructor University Bremen, Germany}
\date{\today}

\begin{abstract}
A recent trend in mathematical modeling is to publish the computer code together with the research findings. Here we explore the formal question, whether and in which sense a computer implementation is distinct from the mathematical model. We argue that, despite the convenience of implemented models, a set of implicit assumptions is perpetuated with the implementation to the extent that even in widely used models the causal link between the (formal) mathematical model and the set of results is no longer certain.
Moreover, code publication is often seen as an important contributor to reproducible research, we suggest that  in some cases the opposite may be true.
A new perspective on this topic stems from the accelerating trend that in some branches of research only implemented models are used, e.g., in artificial intelligence (AI). With the advent of quantum computers we argue that completely novel challenges arise in the distinction between models and implementations.
\end{abstract}

\maketitle


\section{\label{intro}Introduction}

The use of mathematical models~\footnote{In this paper we do not question the different views on what mathematics is such as a  Platonists', formalists', logicists', intuitionists, empiricists' or  structuralists' views, rather we purely are interested in how it is used in today's science.} has become increasingly commonplace in many fields, including biology, physics, economics, sociology and engineering. Mathematical models are abstract representations of systems or processes that allow researchers to study their behavior under different constraints, understand the mechanisms behind a set of observations, and, finally, to make predictions. Those predictions are then typically compared with observation in the real world. Often, the basic models are based on observations from the past.

\textit{Is a model equivalent to its computer implementation?} The short answer is, of course, no. It starts by the well-known fact that any number, such as for instant $1$, can only be represented by a finite amount resources on a computer (float, double). Those limitations are well known and studied and have the awareness of modelers and users. However, a detailed reflection of this matter is mandatory, because: \begin{itemize}
\item[(1)] Reflecting on this distinction between a model and its implementation offers insight into our view on mathematical models, how we represent them and how this changes over time.
\item[(2)] Studying the equivalences (or lack thereof) of mathematical models and their implementations can allow us to understand some core reasons for the lack of reproducibility of computational results  (see, e.g., \cite{boulesteix2010over,milkowski2018replicability,tiwari2021reproducibility}).
\item[(3)] The topic draws attention to recent trends and innovations in academic publishing, which shape future research.
\end{itemize}

Code submission for mathematical or computational models has become a frequent requirement of funding agencies and academic journals alike. However, in the literature discussing this requirement \cite{stodden2013toward,mckiernan2016open,stodden2016enhancing,barton2022make} the differences between model and code highlighted here, as well as the potential disadvantages of resorting solely to implemented models, are hardly discussed.

A model is formulated on the level of equations (or any mathematical language appropriate for the system at hand). It is based on explicit and implicit assumptions and mostly allows for some consistence checks or proofs, for instance a result can be only a positive number and never a negative or complex number. Mathematical models are abstract formulations derived from human thought processes. They come in the form of relations between numbers, in the form of formulas or a set of equations. Often, those equations cannot be studied fully analytically  (i.e., just in its original functional form).  As a consequence, the original description is mathematical, but implementations of such a model are created for the purpose of running computer simulations or evaluating the model beyond what is analytically possible.

An implicit assumption of a 'good' computer implementation is that the results are a consequence of the mathematical model and independent of the model's specific implementation. This also means that any aspect pertaining solely to the computer implementation of the model is irrelevant for the model behavior and for the results obtained with the model.
Examples of such implementation details are the choice of initial conditions, the sequence, in which operations formalized in the model are executed, and the numerical recipes employed for solving or simulating the model.

In some fields of science, the strength of mathematical modeling is to provide precise quantitative predictions of some measurable quantities (see also Fig.~\ref{fig1}). In these cases, the model stems from a theory that claims to represent this aspect of the world with maximal fidelity. The quality of an implementation based on the mathematical model, which in itself is based on some theoretical model, is then typically quantified by the closeness of the predicted quantity to the measurement quantity. Even if successful in this sense, the application to another case may lead to discrepancies, requiring  a change in the implementation or its underlying model.

In other cases a qualitative agreement with observations is targeted, as in the case of modeling socioeconomic systems \cite{castellano2009statistical,pereira2017econophysics}  or other complex systems \cite{holovatch2017complex,barthelemy2019statistical,fan2021statistical}. Then the claim of the modeling effort is that the model contains all the mechanistic ingredients to capture a certain (often counter-intuitive) phenomenon. In these fields of science particular importance is given to the smallest, most minimal models capturing an obverved behavior (smallest in the sense of containing the smallest number of degrees of freedom and/or the smallest number of parameters). Such 'toy models' or minimal models are a cornerstone of many applications of, e.g., statistical physics \cite{kauffman1969metabolic,bak1988self,rodrigues2016kuramoto}.

With this perspective paper we want to draw attention to the non-trivial and crucial relationship between a mathematical model and its computer implementation and the subtle and informative ways, in which they can be different. But we also want to emphasize some potential challenges, which come along with the otherwise commendable trend in academic publishing of requiring code submissions for mathematical models \cite{tiwari2021reproducibility,vieira2022computational}, because we feel that the current debate at times underestimates the slight distortions of the modeling landscape that come along with it.

\section{\label{details} A detailed assessment}

In this section we illustrate our general point about the impact of code availability for mathematical models on model diversity and provide more details about the distinction of models from their implementation.

\subsection{Technical issues in the implementation}

\begin{figure*}[ht!]
\centerline{\includegraphics[width=16cm]{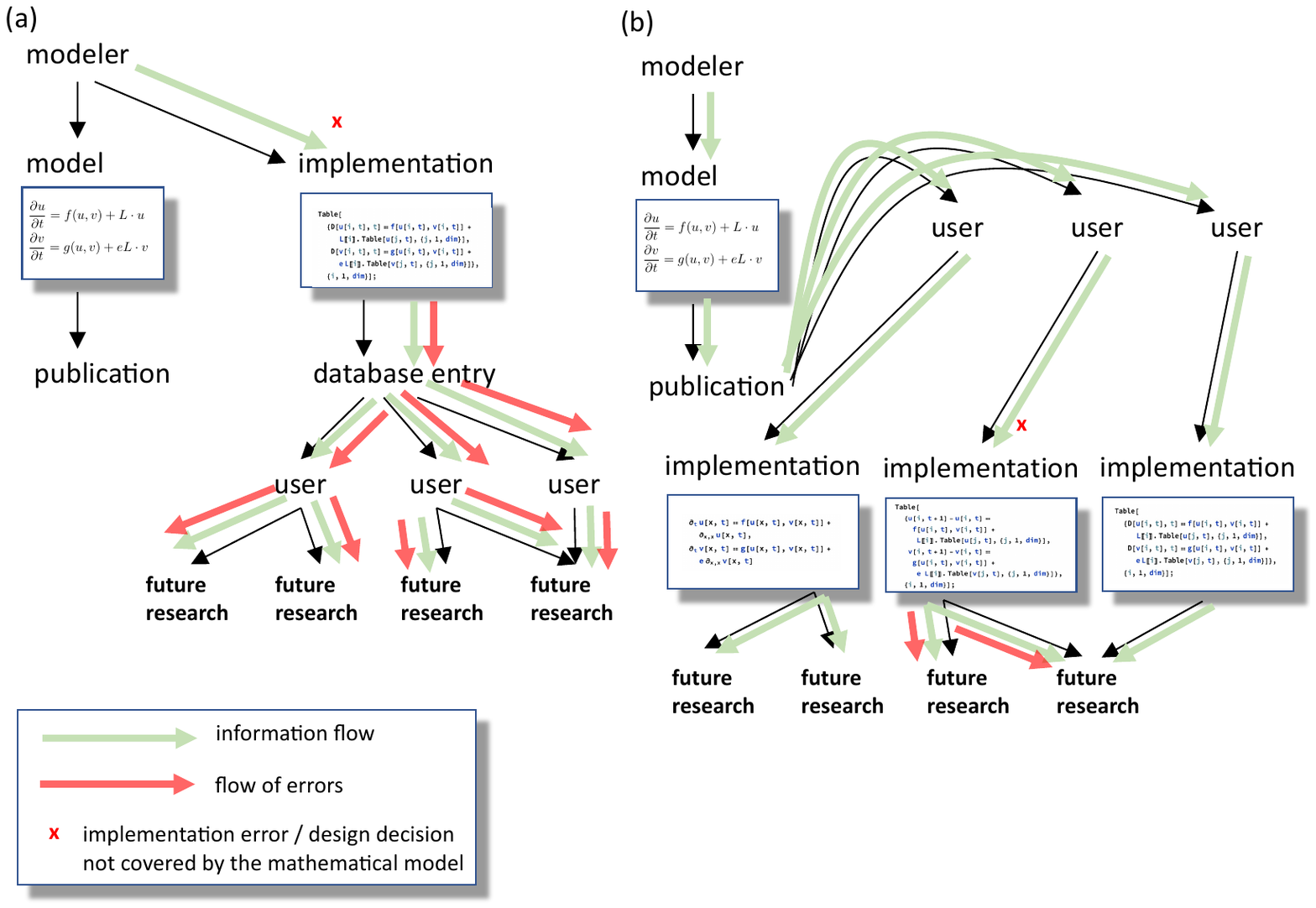}}
\caption{\label{figR1} Schematic representation of the difference between (a) model implementations published along with the academic results and (b) the publication serving as the main source of information about the mathematical model. In (a) information flow and the flow of errors or the impact of design decisions, which are not part of the original mathematical model, are coupled, while in (b) due to the diversity of implementations by different modelers those decisions by a single modeler are not propagated.}
\end{figure*}

Re-implementations decouple the information flow from the flow driven by small errors and implementation design decisions not covered by the original mathematical model. As mentioned before, among these are  discretization effects, choices of initializations, tie-breaking criteria, the order of executing logical steps and many more. Figure \ref{figR1} is a schematic illustration of this change in information flow (with information flow given in green and flow of errors and implementation design decisions given in red).

As an early example of the challenging relationship between a model and its computer implementation, Edward Lorenz's numerical experiments with a $12$-variable version of his weather model in the early 1960s come to mind. The rounding errors of the computer (providing initial conditions only to the third decimal in the output) were enough to lead to drastically different simulation runs \cite{gleick2008chaos}. These observations -- and Lorenz's ingenuity of interpreting them -- culminated in the discovery of a deterministic model represented as coupled ordinary differential equations (ODEs) -- the Lorenz equations -- in his  seminal paper in 1963 \cite{lorenz1963deterministic}.

 Some practical aspects also affect this debate. Model implementations can be platform dependent (using packages, which change over time) or dependent on the numerical recipes (e.g., ODE solvers) used. In the computer algebra and programming environment \textit{Mathematica}, for example, a matrix consisting of entries $1/2$  and a matrix consisting of entries $0.5$ can yield markedly different results in terms of representation, normalization and numerical accuracy, as the first variant triggers the symbolic routines, while the second triggers the numerical routines of this programming environment.

Summarizing, the list, in which a model can differ from its technical implementation, is quite long. It includes factors like
initialization, discretization (e.g., within a solver), implementation errors, statistics (length of the simulation, transient, number of simulations), tiebreaking criteria, order of update rules, noise implementation (and quality of the random number generator) and many more.



\subsection{Reproducibility -- a key argument for the quantity of research?}

\textit{What can be learned from discussing the relationship between a model and its implementation?} The best discussed aspect of this topic is certainly the reproducibility of scientific results. In Ref.~\cite{tiwari2021reproducibility}, for example, published (and electronically available) models have been assessed regarding their capability to produce the results of the core publication behind each model. The striking observation in this field of biological modeling, namely how limited this direct reproducibility is, has led to a range of recommendations regarding the publication of model implementations and the curation of mathematical models. A strong argument in favor of this general procedure is provided by evidence that reproducible models receive more citations \cite{hopfl2023bayesian}.

Other aspects of these topics are general error propagation and the propagation of implicit knowledge beyond the original model equations. As a trivial example, to select one element among the techical issues discussed above, a particular way of choosing initial conditions for a system of coupled differential equations hard-wired into the implementation might prevent a full numerical exploration of the dynamical scope of the model, unless the implementation is altered in this respect or the model is re-implemented. Re-implementing existing mathematical model is also of high educational value and is often a good starting point for a young graduate student. Moreover, it is a kind of quality check between implementation and mathematical model (by which errors have often been identified in the past, even by models widely used.

Note that here our focus is on highlighting a potential danger in this practice of required / recommended code publication. It needs to be emphasized, however, that indeed availability of implemented models has hugely amplified model (re-)usage and comparison of data with models.




\section{\label{casestudies} Beyond standard mathematical models: AI and quantum computers}









\subsection{Data-driven modeling: AI}

Some insight in the relationship between a model and its implementation can be gained by imagining to discard the mathematical model completely and rather directly implement all systemic knowledge in the form of a computer program. It then becomes obvious that, indeed, the formulation of a mathematical model and the design of a computer implementation are also conceptually quite different tasks. If you strive for an implementation only, the choice of functions (e.g., linear, quadratic, sigmoidal, ...) and conciseness may be less relevant. If you strive for a mathematical model, elegance and simplicity are important factors \cite{Farmelo2002}.  Conceptual mistakes are more easily spotted in equations than in their implementations.

Artificial intelligence (AI) is an example of such a purely data-driven approach to modeling. There, the implementation task is fully given to the training process of the device using available data, and the distinction between model and implementation becomes impossible. We all experience currently the experiment how far this approach can be taken and which social-economic developments will be triggered.

\begin{figure*}[ht!]
\centerline{\includegraphics[width=16cm]{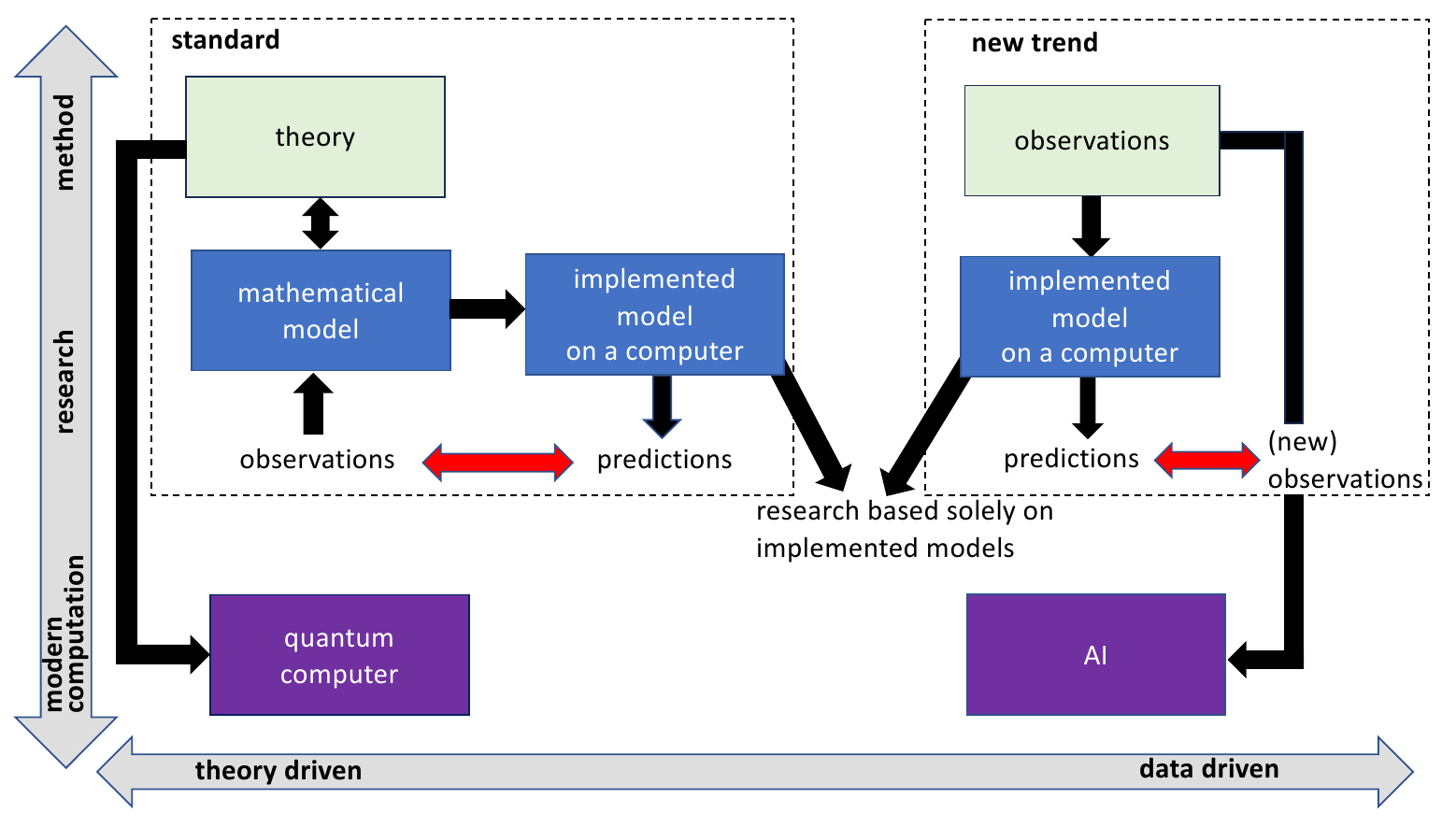}}
\caption{\label{fig1} {Summary of the concepts driving the distinction between models and their implementation. Left: Standard (historical) approach, where theory-driven mathematical models are explored on their own and with the help of computer simulations. This implements a cycle model $\rightarrow $ implementation $\rightarrow $ prediction $\rightarrow $ observations $\rightarrow $ model, which also challenges, and eventually contributes to, theory. The largest impact of this approach on modern computation is in the form of quantum computers.  Right: New trend, where observations are directly translated into implementations (sometimes via a mediating mathematical model, which however stays in the background, because the implementation is regarded as the main outcome of the scientific endeavor). This process creates predictions, which can be compared to new observations. The largest impact of this approach on modern computation is in the form of artificial intelligence (AI).
}}
\end{figure*}

Both avenues of modeling, theory-driven and data-driven, have given rise to modern forms of computation, quantum computers in the theory-driven case, AI in the data-driven case. Figure \ref{fig1} summarizes this situation.

After briefly highlighting the role of AI in this discussion of the distinction between a model and its implementation, we will in the following section discuss the other cornerstone of modern computation, quantum computers.

\subsection{Quantum computers}
Quantum computers challenge the classical distinction between model and implementation in a multitude of ways. In the following we will briefly address this novel form of computation and discuss it in the context of models and their implementation.

With the advent of quantum computers novel computing devices will be on the market that will definitely ``compute'' or transfer information in a very different way than classical computers. Here the implementation is even be much more different since the quantum mechanical theory behind a given algorithm is much deeper than in the classical case. Let us explain this in more details.

Let us consider the easiest case, the encoding of a bit, i.e. ``0'' or ``1''. Classically, we need a system that can be in two alternative states, for which we define one as ``0'' and the other as ``1''. The reading device needs to be able to distinguish the two cases (in the optimal case without any error). In a quantum computer the bit information needs to be encoded in its building blocks, e.g. in a quantum mechanical systems that ``answers'' by a given measurement device --in the easiest case--  with a dichotomic answer, which again will be defined as ``0'' and ``1''.

In strong contrast to classical systems, the states allowed by any such quantum system is given by~\footnote{Without loss of generality we consider only pure states, i.e. states to which no classical uncertainty is added.}
\begin{equation}
    |\psi\rangle=\cos\frac{\theta}{2}\;|\textrm{``0''}\rangle+\sin\frac{\theta}{2}\cdot e^{i \phi}\;|\textrm{``1''}\rangle
\end{equation}
which is called a ``quantum bit'' or shortly ``qubit''. It contains in a sense infinite information since the values $\theta,\phi$ can be taken from an infinite set of numbers ($\theta\in[0,\pi]$ and $\phi\in[ 0,2\pi\}$). Consequently, a quantum state can be initiated in infinite many different states (by  choosing different $\theta,\phi$), whereas a classical computer has either voltage $0$ (state ``0'') or voltage $5$ (state ``1''). However, one needs to measure a quantum state, to obtain information about the initialization. And herewith, the information is a one bit information, either ``0'' or ``1''. Of course, this is the same for classical computer, you need to use e.g. a Voltmeter to measure the voltage.

So, what role do $\theta,\phi$ play? Those reveal, when one repeats the procedure of initialisation and measuring. Then from this (infinite) collection of measurements one will find that in a series of ``0'' occur with probability $\cos^2\frac{\theta}{2}$ and ``1'' with the probability   $\sin^2\frac{\theta}{2}=1-\cos^2\frac{\theta}{2}.$ (The parameter $\phi$ only reveals if the state is rotated by some device or a different measurement than ``0'' and ``1'' is chosen).

So the point is that on the theory level, we have the full information ($\theta,\phi$) since we can and must choose them for the initiation, but, depending on the chosen experimental setup, we get only partial information about the values. In our chosen setup only information about $\theta$ and not $\phi$. Moreover, only if we have a repetition of the process, initialization and measurement, we get successively access to this information, the quantum theory is a probabilistic one.

To sum up, any theoretical model of a quantum mechanical process contains in general more parameters than can be read out in an experiment and at least some repetition, during which one has to imply the hypothesis that no processes take place which would change the observable of interest, is needed to access experimental results. Nowadays, the main issue in the simulation of a quantum process on a classical computer is that a classical computer cannot easily process all parameters, thus short cuts have to be implemented in addition to all the problems listed in the previous sections.

A striking other problem in simulation quantum processes comes with entanglement, which has no classical counterpart. It needs in a classical simulation that parts of computations do depend on each other in a not neccessarily causal way.

Running a quantum algorithm on a quantum computer, on the other hand, solves this problem, however, our current quantum devices struggle with experimental errors (decoherence), for which we so far have not found any simple way to correct for.

\section{\label{conclusion}Conclusion}

In most cases, computer implementations of mathematical models are necessary to actually use those models for prediction or simulation. Therefore, while mathematical models and their computer implementations are technically distinct objects, they are functionally intertwined in many applications. The use of software to solve mathematical models allows for a more practical usage, enabling higher model complexity real-world predictions and applications. Despite their functional interdependence, it is important to understand the distinctions between mathematical models and their computer implementations. Mathematical models are conceptual and theoretical frameworks that describe the behavior of systems, while computer implementations are tangible tools used to simulate or predict system behavior based on those theoretical frameworks.

As so eloquently summarized in \cite{may2004uses}:
\begin{quote}
Until only a decade or two ago, anyone pursuing this kind of activity [numerical simulations of mathematical models] had to have a solid grounding in mathematics. And that meant that such studies were done by people who had some idea, at an intuitive level, of how the original assumptions related to the emerging graphical display or other conclusions on their computer.
\end{quote}
We believe that this process is drastically accelerated (and its implicit dangers are substantially amplified) by the 'implementation monoculture' enabled by the distribution of readily implemented mathematical models.

Implementations of the same mathematical model are not identical.
And from our perspective, understanding is mediated by analyzing equations (often with the help of numerical simulations), not just by running the computer code.

We acknowledge the usefulness of model databases (like BioModels, \cite{le2006biomodels}), in particular in allowing experimentalists the means of directly comparing their experimental findings with available mathematical models. We feel, though, that there is a serious danger in the resulting 'implementation monoculture' and we suggest to incentivize the reimplementation of existing mathematical models, in order to help distinguish between universal results of a mathematical model and those requiring additional implicit ingredients included in a given implementation.

\begin{acknowledgments}

The starting point of this paper is a discussion in July 2022 in Salzburg at the Festive Session 2022 of the European Academy of Sciences and Arts.  B.C.H. acknowledges gratefully that this research was funded in whole, or in part, by the  Austrian Science Fund (FWF) project P36102-N. M.T.H. thanks the Hamburg Institute for Advanced Study (HIAS) for hospitality and the Joachim Herz Stiftung for funding his fellowship at HIAS, during which part of this work has been performed.

\end{acknowledgments}




%

\end{document}